\documentclass[12pt]{article}

\usepackage[margin=1in]{geometry}

\usepackage{setspace}

\usepackage{graphicx}

\usepackage{enumitem}

\usepackage{microtype}
\usepackage{subfigure}
\usepackage{booktabs}
\usepackage{comment}
\usepackage{wrapfig}
\usepackage{arydshln}
\usepackage{enumitem}
\usepackage{multirow}
\usepackage{url}
\usepackage{natbib}
\usepackage{authblk}

\RequirePackage[colorlinks,citecolor=blue,linkcolor=blue,urlcolor=blue,pagebackref]{hyperref}

\usepackage{amsmath}
\usepackage{amssymb}
\usepackage{mathtools}
\usepackage{amsfonts}
\usepackage{bm}
\usepackage{bbm}

\usepackage{algorithm}
\usepackage{algorithmic}

\def\eqref#1{equation~\ref{#1}}

\def\1{\bm{1}}

\DeclareMathAlphabet{\mathsfit}{\encodingdefault}{\sfdefault}{m}{sl}
\SetMathAlphabet{\mathsfit}{bold}{\encodingdefault}{\sfdefault}{bx}{n}

\def\calF{{\mathcal{F}}}

\def\calX{{\mathcal{X}}}

\def\bbE{{\mathbb{E}}}

\def\bbR{{\mathbb{R}}}

\newcommand{\p}[1]{\left(#1\right)}
\newcommand{\sqb}[1]{\left[#1\right]}
\newcommand{\cb}[1]{\left\{#1\right\}}

\newcommand{\bigp}[1]{\big(#1\big)}
\newcommand{\bigsqb}[1]{\big[#1\big]}

\newcommand{\Bigp}[1]{\Big(#1\Big)}

\usepackage{amsthm}

\theoremstyle{plain}

\usepackage[textsize=tiny]{todonotes}
\usepackage{multirow}
\usepackage{wrapfig}
\usepackage{subfigure}

\renewcommand{\eqref}[1]{(\ref{#1})}

\usepackage{xcolor}
\newcount\Comments  
\newcommand{\kibitz}[2]{\ifnum\Comments=1\textcolor{#1}{#2}\fi}

\usepackage{listings}

\newcommand{\ARW}{\mathrm{ARW}}
\newcommand{\RW}{\mathrm{RW}}
\newcommand{\RA}{\mathrm{RA}}

\usepackage{mathtools}
\mathtoolsset{showonlyrefs=true}

\usepackage[capitalize,noabbrev]{cleveref}

\allowdisplaybreaks

\title{Covariate Balancing and Riesz Regression Should Be Guided by the Neyman Orthogonal Score in Debiased Machine Learning}
\author{Masahiro Kato\thanks{Email: \texttt{mkato-csecon@g.ecc.u-tokyo.ac.jp}}$\,$}

\affil{Data Analytics Department, Mizuho-DL Financial Technology, Co., Ltd.}

\date{\today}

\begin{document}

\maketitle 

\begin{abstract}
    This position paper argues that, in debiased machine learning, balancing functions should be derived from the Neyman orthogonal score, not chosen only as functions of covariates. Covariate balancing is effective when the regression error entering the score can be represented by functions of covariates alone, and it is the natural finite-dimensional approximation for targets such as ATT counterfactual means. For ATE estimation under treatment effect heterogeneity, however, the score error generally contains treatment-specific components because the outcome regression is a function of the full regressor $X=\p{D,Z}$. In that case, balancing common functions of $Z$ can leave the treatment-specific component unbalanced. We therefore advocate regressor balancing, implemented by Riesz regression with basis functions of $X$, as the general balancing principle for DML. The position is not that covariate balancing is invalid, but that covariate balancing should be understood as the special case that is appropriate when the score-relevant regression error is a function of covariates alone.
\end{abstract}


\section{Introduction}
In DML, the balancing functions should be chosen from the regression error that appears in the Neyman orthogonal score. In general, this leads to regressor balancing; covariate balancing is the restricted case in which the balanced functions depend only on covariates.

Covariate balancing and debiased machine learning (DML) are widely used in observational causal inference \citep{Chernozhukov2018doubledebiased,Hainmueller2012entropybalancing,Imai2013covariatebalancing}. In this study, we reconsider covariate balancing as a finite-dimensional approximation to the balancing condition induced by the Neyman orthogonal score. This viewpoint shows why covariate balancing is effective when the score-relevant regression error can be represented by functions of covariates alone, and why it can be restrictive when treatment effect heterogeneity makes the relevant error depend on the full regressor $X=\p{D,Z}$.

Covariate balancing methods estimate the propensity score or construct balancing weights by imposing balance restrictions on functions of covariates \citep{Hainmueller2012entropybalancing,Imai2013covariatebalancing,Zubizarreta2015stableweights}. DML, in contrast, starts from a Neyman orthogonal score for the target estimand \citep{Chernozhukov2018doubledebiased,Chernozhukov2024appliedcausal}. These two perspectives are often treated as separate. We argue that they should be connected through the score error: the functions to be balanced should be those that approximate the regression error appearing in the Neyman orthogonal score.

We reconsider covariate balancing from the viewpoint of the Neyman orthogonal score. Although the form of the score is often known, it depends on unknown nuisance parameters. Therefore, we estimate the score by replacing the unknown nuisance parameters with their estimators. From this viewpoint, desirable nuisance parameter estimators are those that reduce the error between the score using the true nuisance parameters and the score using estimated nuisance parameters. In this context, covariate balancing plays an important role because it can eliminate the part of this error that is represented by functions of covariates alone.

However, when the treatment effect is heterogeneous, this cancellation does not generally occur. In such cases, the outcome regression is a function of both treatment and covariates. Therefore, it is more desirable to approximate the Riesz representer using basis functions $\Phi(D,Z)$ that depend on both treatment and covariates. When such basis functions are used and the corresponding balancing condition holds, the part of the score error represented by those basis functions vanishes. This is the basic reason why we focus on regressor balancing.

\subsection{Example: ATE Estimation}
As a running example, we consider average treatment effect (ATE) estimation with observations $\cb{\p{D_i, Z_i, Y_i}}^n_{i=1}$ \citep{Imbens2015causalinference}. Our goal is to estimate the ATE defined as $\theta^{\text{ATE}}_0 \coloneqq \bbE\sqb{\gamma_0(1, Z_i) - \gamma_0(0, Z_i)}$, where $\gamma_0(d, z) \coloneqq \bbE\sqb{Y_i\mid D_i = d, Z_i =z}$ is the regression function. In this problem, the Neyman orthogonal score is given as
\[\psi^{\text{ATE}}\p{D_i, Z_i, Y_i; \alpha_0, \gamma_0, \theta^{\text{ATE}}_0} \coloneqq \alpha_0(D_i, Z_i)\bigp{Y_i - \gamma_0(D_i, Z_i)} + \gamma_0(1, Z_i) - \gamma_0(0, Z_i) - \theta^{\text{ATE}}_0,\]
where $\alpha_0(d, z) \coloneqq \frac{\mathbbm{1}\p{d=1}}{e_0(z)} - \frac{\mathbbm{1}\p{d=0}}{1 - e_0(z)}$ is the Riesz representer, and $e_0(z) \coloneqq \Pr\p{D = 1\mid Z = z}$ is the propensity score. By replacing the unknown $\alpha_0$ and $\gamma_0$ with their estimators $\widehat\alpha$ and $\widehat\gamma$ and solving the estimation equation $\frac{1}{n}\sum^n_{i=1}\psi^{\text{ATE}}\p{D_i,Z_i,Y_i;\widehat\alpha,\widehat\gamma,\theta}=0$ for $\theta$, we obtain an estimator $\widehat\theta$ of the ATE. Let $\xi(d,z)\coloneqq \gamma_0(d,z)-\widehat\gamma(d,z)$. The part of the plug-in score error that depends on $\xi$ is
$
\frac{1}{n}\sum^n_{i=1}
\Bigp{
\widehat\alpha(D_i,Z_i)\xi(D_i,Z_i)-\xi(1,Z_i)+\xi(0,Z_i)
}$. 
If $\xi(d,z)$ is represented by basis functions $\Phi(d,z)$ and
$
\frac{1}{n}\sum^n_{i=1}\widehat\alpha(D_i,Z_i)\Phi(D_i,Z_i)
=
\frac{1}{n}\sum^n_{i=1}\p{\Phi(1,Z_i)-\Phi(0,Z_i)}
$, 
then this deterministic component vanishes for the represented part of $\xi$. This is the basic role of regressor balancing in ATE estimation.

Now consider covariate balancing, where the basis depends only on $Z$; that is, $\Phi(d, z) = \widetilde{\Phi}(z)$. The covariate balancing condition is $\frac{1}{n}\sum^n_{i=1}\widehat{\alpha}(D_i, Z_i) \widetilde{\Phi}(Z_i) = \frac{1}{n}\sum^n_{i=1}\p{\widetilde{\Phi}(Z_i) - \widetilde{\Phi}(Z_i)} = 0$. In contrast, reducing the error between the scores requires that $\xi(d, z) = \gamma_0(d, z) - \widehat{\gamma}(d, z)$ belong to the linear space spanned by $\widetilde{\Phi}(z)$.

Comparing the two cases, we can see that covariate balancing is more restrictive than regressor balancing from the viewpoint of error minimization for the true Neyman orthogonal scores. Covariate balancing works in some special cases, such as when the treatment effect is homogeneous. In more general cases, regressor balancing is more desirable.

\subsection{Our Position and Contribution}
Our position is that, for DML, the relevant balancing condition should be derived from the Neyman orthogonal score. This condition is regressor balancing when the score-relevant regression error is a function of the full regressor $X$, and it reduces to covariate balancing when that error is a function of $Z$ alone. Thus, covariate balancing is not invalid; it is a special case whose appropriateness depends on the target estimand and the regression component entering the score error. Our contribution is to make this distinction explicit and to organize existing balancing methods through the functions they balance and the way they control weight stability.

Our study builds on several lines of work in causal inference. Riesz representer estimation based on the imbalance $\Delta_n(\widehat\alpha,f)$ has been studied in \citet{Chen2014sievem,Chen2015sievewald}. Squared error minimization type Riesz representer estimation is proposed in \citet{Chernozhukov2021automaticdebiased}, and \citet{BrunsSmith2025augmentedbalancing} shows that augmented balancing weights and regression are closely related in linear spaces. Stable balancing weights directly control covariate balance and weight stability \citep{Zubizarreta2015stableweights}. Tailored loss methods show that propensity score estimation can be designed to improve covariate balance for the target estimand \citep{Zhao2019covariatebalancing}. Entropy balancing directly constructs weights that match prespecified covariate moments \citep{Hainmueller2012entropybalancing}. \citet{Kato2025directbias} points out that Riesz regression can be derived from the viewpoint of density ratio estimation (DRE), and \citet{Kato2026aunified} unifies the existing approaches. These works suggest that balancing can reduce the error for the true Neyman orthogonal score. We formally state this relationship and clarify its implications for the choice of balancing functions.

\paragraph{Contents.}
Section \ref{sec:setup} formulates the problem within the DML framework. In Section~\ref{sec:estimators}, we introduce candidates of estimators, discuss asymptotic efficiency, and raise issues about finite sample performance. In Section~\ref{sec:neymanerrorminimization}, we show that regressor balancing minimizes the error for the true Neyman orthogonal score. In Section~\ref{sec:riesz_regression}, we explain why Riesz regression automatically induces regressor balancing when basis functions are used. In Section~\ref{sec:ate_estimation}, we reconsider covariate balancing in ATE estimation. In Section~\ref{sec:att_estimation}, we explain why covariate balancing can be sufficient in ATT estimation. In Section~\ref{sec:discussion}, we discuss related estimators and existing work. Section~\ref{sec:experiment} discusses experiments, and the last section concludes.

\section{Setup}
\label{sec:setup}

We introduce the general setup for DML. Let $W=(X,Y)$ be an observation, where $Y$ is an outcome and $X$ is a regressor. In treatment effect problems, we often write $X=\p{D,Z}$, where $D$ is a treatment indicator and $Z$ is a vector of covariates. Let $P_X(A)\coloneqq\Pr\p{X\in A}$ denote the marginal distribution of $X$. Define $L_2(P_X)\coloneqq\cb{f:\calX\to\bbR, f\text{ measurable}:\int_{\calX} f(x)^2dP_X(x)<\infty}$, with functions identified if they are equal $P_X$-almost surely. Let $\gamma_0(x)=\bbE\sqb{Y\mid X=x}$ be the regression function.

\paragraph{Observations and estimand.}
Suppose that we observe i.i.d. observations $\cb{W_i}^n_{i=1}$ from the distribution of $W$. We consider an estimand $\theta_0$ characterized as a linear functional of $\gamma_0$:
\[\theta_0 \coloneqq \bbE\sqb{m(W; \gamma_0)},\]
where $m(W; \gamma)$ is a known functional that is linear in $\gamma$. By the Riesz representation theorem, there exists $\alpha_0\in L_2(P_X)$ such that
\[\bbE\sqb{m(W;\gamma)}=\bbE\bigsqb{\alpha_0(X)\gamma(X)}
  \quad\text{for all } \gamma\in L_2(P_X).
\]
We call $\alpha_0$ the Riesz representer. In DML, the corresponding Neyman orthogonal score is
\[\psi(W;\eta_0,\theta_0)\coloneqq \alpha_0(X)\bigp{Y-\gamma_0(X)}+m(W;\gamma_0)-\theta_0,\]
where $\eta_0=(\alpha_0,\gamma_0)$.

\paragraph{Examples.}
We can derive various estimands by specifying the functional $m$. Examples include ATE and ATT counterfactual mean as follows:
\begin{itemize}[noitemsep, topsep=0pt, leftmargin=0.40cm]
    \item ATE: $m(W; \gamma) = \gamma(1, Z) - \gamma(0, Z)$.
    \item ATT counterfactual mean: $m(W; \gamma) = \frac{D}{\Pr(D=1)}\gamma(0,Z)$.
\end{itemize}
Functionals involving derivatives, such as AME, fit the same logic after replacing the $L_2(P_X)$ domain with an appropriate smoothness class.

\paragraph{Our goal.}
Our goal is to construct estimators of $\theta_0$ with desirable asymptotic and finite sample properties. We use two criteria to judge the soundness of the estimator. The first criterion is asymptotic efficiency in the sense of semiparametric efficiency theory, that is, the asymptotic variance of the estimator matches the efficiency bound, the theoretically best asymptotic variance among regular estimators. See \citet{VanderVaart1998asymptoticstatistics}. The second criterion is finite sample performance. In many cases, asymptotic efficiency is theoretically investigated, while finite sample performance is empirically evaluated.

\section{Estimators, Asymptotic Efficiency, and Finite-Sample Issues}
\label{sec:estimators}
This section introduces candidate estimators and discusses their properties. We first use asymptotic efficiency as the benchmark. DML provides a way to construct asymptotically efficient estimators, but asymptotic efficiency does not by itself guarantee finite-sample performance. This motivates the finite-sample perspective based on regressor balancing. The main point is that DML and covariate balancing serve different roles. From the viewpoint of asymptotic efficiency, the ARW estimator is optimal in the sense that its asymptotic variance matches the efficiency bound. However, asymptotic optimality does not settle questions about finite-sample performance.

\subsection{Candidates of Estimators}
In this study, we consider the following three types of estimators: the Augmented Riesz Weighting estimator
\[\widehat\theta^{\ARW}
  =\frac{1}{n}\sum^n_{i=1}\p{m(W_i;\widehat\gamma)+\widehat\alpha(X_i)(Y_i-\widehat\gamma(X_i))};\]
the Riesz Weighting (RW) estimator $\widehat\theta^{\RW}=\frac{1}{n}\sum^n_{i=1}\widehat\alpha(X_i)Y_i$, and the Regression Adjustment (RA) estimator $\widehat\theta^{\RA} =\frac{1}{n}\sum^n_{i=1}m(W_i;\widehat\gamma)$.

As we explain in the subsequent subsections, the ARW estimator is closely connected to the semiparametric efficiency bound and Neyman orthogonal scores. The RW estimator is a generalization of the IPW estimator in ATE estimation. The RA estimator is a plug-in estimator based only on the regression function. These estimators differ in which nuisance parameter they use and in how they respond to finite sample errors in nuisance estimation.

\subsection{Asymptotic Efficiency and DML}
We consider three types of estimators as candidates. The next question is which estimator is preferable among them. To address this question, we introduce asymptotic efficiency theory.

\paragraph{Asymptotic efficiency bound.}
We consider the efficiency bound, called the Le Cam-Hajek bound \citep{LeCam1986asymptoticmethods} or the semiparametric efficiency bound \citep{VanderVaart1998asymptoticstatistics}. The asymptotic efficiency bound gives the theoretically best asymptotic variance among regular estimators. For details, see \citet{Bickel1998efficientand} and \citet{VanderVaart1998asymptoticstatistics}.

\paragraph{Regular and asymptotically linear (RAL) estimators.}
An estimator whose asymptotic variance matches the efficiency bound is called asymptotically efficient. It is known that an RAL estimator with the efficient influence function is asymptotically efficient. RAL estimators can be written as $\sqrt{n}\p{\widehat{\theta} - \theta_0} = \frac{1}{\sqrt{n}}\sum^n_{i=1}\psi(W_i; \eta_0, \theta_0) + o_p(1)$ ($n\to \infty$), 
where $\psi$ is the efficient influence function depending on some nuisance parameter $\eta_0$ and the estimand $\theta_0$.

\paragraph{Neyman orthogonal score and Riesz representer.}
In the DML framework, we estimate the estimand by solving the estimation equation with the Neyman orthogonal score and estimated nuisance parameters. Consider the case where we replace the nuisance parameter $\eta_0$ in the efficient influence function with its estimator $\widehat{\eta}$.

If the first order effect of nuisance estimation error vanishes, such an efficient influence function is called a Neyman orthogonal score. Although the efficient score and efficient influence function are sometimes defined differently, we use the same object for simplicity. The Neyman orthogonal score is defined for each estimand and usually takes the following form:
\[\psi(W_i; \eta_0, \theta_0) \coloneqq \alpha_0(X_i)\bigp{Y_i - \gamma_0(X_i)} + m(W_i;\gamma_0) - \theta_0,\]
where $\alpha_0 \in L_2(P_X)$ is the Riesz representer, and $\eta_0 \coloneqq (\alpha_0, \gamma_0)$ is the nuisance parameter, a set of the Riesz representer $\alpha_0$ and the regression function $\gamma_0$.

The Riesz representer is given from the following Riesz representation theorem:
\[\bbE\sqb{m(W;\gamma)}=\bbE\bigsqb{\alpha_0(X)\gamma(X)}
  \quad\text{for all } \gamma\in L_2(P_X).\]

Under the true nuisance parameter $\eta_0$, $\bbE\sqb{\psi(W;\eta_0,\theta_0)}=0$ holds. By replacing the unknown nuisance parameter $\eta_0$ with its estimator $\widehat{\eta}$ and expectation with the sample mean, we estimate $\theta_0$ by solving the estimation equation $\frac{1}{n}\sum^n_{i=1}\psi(W_i; \widehat{\eta}, \theta) = 0$ for $\theta$.

\paragraph{Asymptotic efficiency of the ARW estimator.}
The solution of the above estimation equation is the ARW estimator. Under standard DML conditions, including suitable convergence rates for the nuisance estimators and either empirical process restrictions or cross fitting, the ARW estimator is asymptotically normal and efficient \citep{Chernozhukov2018doubledebiased}. Our focus is different: even when asymptotic efficiency is guaranteed, finite sample behavior still depends on the empirical error of the estimated score.

\subsection{Issues of Finite-Sample Performance}
Thus, the ARW estimator with suitably constructed nuisance estimators is asymptotically normal and efficient. Therefore, in the asymptotic regime, no regular estimator has a smaller asymptotic variance under the same model. However, this asymptotic optimality does not guarantee finite sample performance. In a finite sample, the estimator can still be sensitive to the quality of the estimated score. Covariate balancing is related to this finite sample concern because it can reduce some components of the score error. The next section formalizes this connection through the Neyman error.

\section{Neyman Error Minimization via Regressor Balancing}
\label{sec:neymanerrorminimization}
We consider estimators $\widehat{\theta}$ with the form $\frac{1}{n}\sum^n_{i=1}\psi\p{W_i; \widehat{\gamma}, \widehat{\alpha}, \widehat{\theta}} = 0$. For general $\widehat{\gamma}, \widehat{\alpha}$, we obtain the ARW estimator, while as special cases with $\widehat{\gamma}(x) = 0$ and $\widehat{\alpha}(x) = 0$, we obtain the RW and RA estimators, respectively.

\subsection{Neyman Error}
As discussed above, an ideal estimator would solve $\frac{1}{n}\sum^n_{i=1}\psi\p{W_i;\gamma_0,\alpha_0,\theta}=0$ for $\theta$. The plug-in score replaces the true score with its estimated counterpart. We focus on the part of the plug-in score error that is affected by the estimated nuisance functions, and define
\begin{align}
  \mathrm{NE}_n(\widehat\gamma,\widehat\alpha)
  \coloneqq
  \frac{1}{n}\sum^n_{i=1}
  \Bigp{
  \widehat\alpha(X_i)\bigp{Y_i-\widehat\gamma(X_i)}
  +m(W_i;\widehat\gamma)-m(W_i;\gamma_0)
  }.
\label{eq:neyman_error}
\end{align}
We call this quantity the Neyman error. It measures the empirical discrepancy in the estimated score after removing the target component $m(W_i;\gamma_0)$.

\subsection{Regressor Balancing}
This section defines regressor balancing formally. For a candidate representer $\alpha$ and a function $f \in \calF$, we define a balancing gap as
\begin{align}
    \Delta_n(\alpha,f)
  \coloneqq\frac{1}{n}\sum^n_{i=1}\Bigp{\alpha(X_i)f(X_i)-m(W_i;f)}.
\end{align}
A representer estimate $\widehat\alpha$ is said to exactly balance a class $\calF$ if $\Delta_n(\widehat\alpha,f)=0$ for all $f\in\calF$, and to approximately balance $\calF$ at tolerance $\delta_n$ if $\sup_{f\in\calF} |\Delta_n(\widehat\alpha,f)|\leq \delta_n$.

Regressor balancing refers to balancing functions $f(X)$ in a function class for the full regressor $X$. When $X=\p{D,Z}$, these functions may depend on both treatment and covariates. Covariate balancing is the restricted case in which $\calF$ contains only functions of $Z$.

\subsection{Neyman Error Minimization via Regressor Balancing}
Using linearity of $m$, it holds that
\begin{align}
  \mathrm{NE}_n(\widehat\gamma,\widehat\alpha)
  &=\frac{1}{n}\sum^n_{i=1}\widehat\alpha(X_i)\varepsilon_i
    -\Delta_n\p{\widehat\alpha,\widehat\gamma-\gamma_0}
\label{eq:neyman_decomp}
\end{align}
where $\varepsilon_i=Y_i-\gamma_0(X_i)$.
If the observations used to evaluate the score are independent of the data used to construct $\widehat\alpha$, then, conditional on the training data and on $X_1,\ldots,X_n$, the first term is a mean-zero weighted noise term. Without such sample separation, this term is still part of the empirical score error, but it should not be described as conditionally mean zero. The second term $-\Delta_n\p{\widehat\alpha,\widehat\gamma-\gamma_0}$ is the deterministic drift induced by regressor imbalance. 

This decomposition gives the main reason for regressor balancing. If $\widehat\gamma-\gamma_0$ belongs to $\calF$ and $\widehat\alpha$ exactly balances $\calF$, then the deterministic drift vanishes. More generally, if $\widehat\gamma(X)-\gamma_0(X)=f(X)+r(X)$ holds for all $f\in\calF$, 
then exact balancing of $\calF$ leaves only $\Delta_n(\widehat\alpha,r)$. Thus, regressor balancing reduces the score error only to the extent that the balanced functions approximate the score-relevant regression error.

\section{Riesz Regression as Automatic Regressor Balancing and Automatic Neyman Error Minimization with Approximation using Basis Functions}
\label{sec:riesz_regression}
This section explains how Riesz regression implements regressor balancing when the Riesz representer and the regression error are approximated by basis functions. The main point is simple. Riesz regression estimates the Riesz representer by solving an empirical risk minimization problem. The first-order condition of this problem gives balancing equations for the basis functions. Therefore, Riesz regression is not only an estimator of the Riesz representer. It is also a way to construct weights that reduce the deterministic part of the Neyman error.

\subsection{Series Estimation of Riesz Representer and Regression Function}
Suppose that we use a vector of basis functions $\Phi(X) = (\Phi_1(X),\ldots,\Phi_p(X))^\top$. We approximate the Riesz representer by a linear model $\alpha_\beta(X)=\beta^\top\Phi(X)$. The regression function, or the regression error $\widehat\gamma-\gamma_0$, is also approximated by the same basis functions or by a related basis. This series approximation is useful because the balancing condition can be checked component by component.

For example, in Riesz regression under the squared loss, we estimate $\beta$ by minimizing $\frac{1}{n}\sum^n_{i=1}\alpha_\beta(X_i)^2
-\frac{2}{n}\sum^n_{i=1}m(W_i;\alpha_\beta)
+\lambda J(\beta)$, 
where $\lambda J(\beta)$ is a regularization term with $\lambda\geq0$. If $\widehat\beta$ is a minimizer, $\widehat\alpha(X)=\alpha_{\widehat\beta}(X)$, and $J$ is differentiable at $\widehat\beta$, then the first-order condition for the $j$th coefficient gives $0
=
\frac{2}{n}\sum^n_{i=1}\Phi_j(X_i)\widehat\alpha(X_i)
-\frac{2}{n}\sum^n_{i=1}m(W_i;\Phi_j)
+\lambda \partial_j J(\widehat\beta)$. 
Equivalently, we have $\Delta_n(\widehat\alpha,\Phi_j)
=
-\frac{\lambda}{2}\partial_j J(\widehat\beta)$. 
Therefore, when $\lambda=0$, Riesz regression exactly balances the basis functions $\Phi_j$. When $\lambda>0$, the remaining imbalance is determined by the regularization term.

\subsection{Regressor Balancing and Neyman Error}
The preceding first-order condition explains why Riesz regression is useful for controlling the deterministic part of the Neyman error. Suppose that the regression error can be decomposed as $\widehat\gamma(X)-\gamma_0(X)
=
\rho^\top\Phi(X)+r(X)$, 
where $r$ is an approximation error. By linearity, we have $\Delta_n\p{\widehat\alpha,\widehat\gamma-\gamma_0}
=
\sum^p_{j=1}\rho_j\Delta_n(\widehat\alpha,\Phi_j)
+
\Delta_n(\widehat\alpha,r)$. 
Thus, balancing the basis functions $\Phi_j$ controls the represented part of the deterministic drift, while the remaining term is determined by the approximation error.

This is the sense in which Riesz regression gives automatic regressor balancing. The balancing condition is not imposed after estimating $\widehat\alpha$. It appears as the first-order condition of the Riesz regression problem. More general versions of Riesz regression, including generalized Riesz regression based on other loss functions and link functions, also yield balancing equations under suitable specifications \citep{Kato2026aunified}. 

\section{ATE Estimation and Reconsidering Covariate Balancing}
\label{sec:ate_estimation}
We now return to ATE estimation and reconsider covariate balancing. In ATE estimation, $X=\p{D,Z}$ and $m(W;\gamma)=\gamma(1,Z)-\gamma(0,Z)$. Therefore, for a generic function $f(d,z)$, the regressor balancing condition is $\frac{1}{n}\sum^n_{i=1}\widehat\alpha(D_i,Z_i)f(D_i,Z_i)
=
\frac{1}{n}\sum^n_{i=1}\p{f(1,Z_i)-f(0,Z_i)}$. 
This condition is defined for functions of the full regressor $X=\p{D,Z}$.

Covariate balancing is obtained by restricting $f(d,z)$ to functions that do not depend on $d$. Let $f(d,z)=h(z)$. Then $m(W;f)=h(Z)-h(Z)=0$, and the balancing condition becomes $\frac{1}{n}\sum^n_{i=1}\widehat\alpha(D_i,Z_i)h(Z_i)=0$. 
For the ATE Riesz representer, this is the usual signed balance between treated and control groups after weighting. Thus, covariate balancing is a restricted case of regressor balancing in ATE estimation.

This restriction is sufficient when the relevant regression error is represented by functions of $Z$ alone. To see the limitation, write the score error as $\xi(D,Z)=\xi_0(Z)+D\xi_1(Z)$. The deterministic part of the ATE score error is $\frac{1}{n}\sum^n_{i=1}\widehat\alpha(D_i,Z_i)\xi(D_i,Z_i)
-
\frac{1}{n}\sum^n_{i=1}\p{\xi(1,Z_i)-\xi(0,Z_i)} =\frac{1}{n}\sum^n_{i=1}\widehat\alpha(D_i,Z_i)\xi_0(Z_i)
+
\frac{1}{n}\sum^n_{i=1}\widehat\alpha(D_i,Z_i)D_i\xi_1(Z_i)
-
\frac{1}{n}\sum^n_{i=1}\xi_1(Z_i)$. 
The first term on the right-hand side is controlled by covariate balancing if $\xi_0$ is in the balanced covariate class. The remaining two terms involve the treatment-dependent component $D\xi_1(Z)$. They generally require a balancing condition for functions that depend on both $D$ and $Z$. Thus, under treatment effect heterogeneity, the outcome regression generally depends on the full regressor and covariate balancing can be restrictive.

This point is closely related to optimal CBPS. \citet{Fan2021optimalcovariate} considers the decomposition $\bbE\sqb{Y(0)\mid Z}=K(Z)$ and $\bbE\sqb{Y(1)-Y(0)\mid Z}=L(Z)$, and shows that the optimal balancing functions for ATE estimation depend on both $K$ and $L$. In particular, the component related to $L$ is needed when treatment effects are heterogeneous. This result is consistent with our position that the balancing functions should be chosen from the error structure of the Neyman orthogonal score, not only from covariates themselves.

\section{ATT Estimation and Sufficiency of Covariate Balancing}
\label{sec:att_estimation}
This study does not deny covariate balancing. Rather, it clarifies when covariate balancing is the relevant balancing condition. One important case is ATT counterfactual mean estimation, where the score-relevant regression component is a function of $Z$ alone and covariate balancing can be sufficient.

In ATT estimation, the relevant component of the estimand can be reduced to the counterfactual mean of the untreated potential outcome among treated units. For this estimation, we use $\gamma_0(0,Z)$ and do not use $\gamma_0(1,Z)$. Here, $\gamma_0(0,Z)$ is a function of $Z$ alone. Therefore, if $\gamma_0(0,Z)$ is well approximated by basis functions of $Z$, then balancing those basis functions between the treated group and the weighted control group directly targets the relevant regression component.

This is the main reason why entropy balancing is natural for ATT-type targets. \citet{Hainmueller2012entropybalancing} proposes entropy balancing as a method for computing weights so that the reweighted control group and the treated group satisfy prespecified moment conditions. In ATT estimation, assigning weights to the control group so that its covariate moments match those of the treated group is a direct finite-dimensional approximation to the counterfactual mean problem.

Thus, the distinction between covariate balancing and regressor balancing depends on the target estimand. For ATT estimation or counterfactual mean estimation in that task, we only use the regression function $\gamma_0(0,Z)$, which can be regarded as a function of $Z$ alone. In contrast, for ATE under treatment effect heterogeneity, the relevant regression function is generally a function of $X=\p{D,Z}$. This is why covariate balancing can be sufficient in the former case but restrictive in the latter case.

\section{Discussion}
\label{sec:discussion}
In this section, we discuss related topics. Also see Appendix~\ref{sec:alternative_views}. 

\paragraph{RW estimator.}
The RW estimator uses only the estimated Riesz representer and the outcome. Therefore, if the estimated Riesz representer satisfies strong balancing conditions for the relevant regression components, the RW estimator can behave like an estimator based on an orthogonal score. This is because exact balancing can replace the missing regression term in the estimating equation. More precisely, if $\Delta_n(\widehat\alpha,\gamma_0)=0$, then we have $\frac{1}{n}\sum^n_{i=1}\widehat\alpha(X_i)Y_i
=
\frac{1}{n}\sum^n_{i=1}\p{m(W_i;\gamma_0)+\widehat\alpha(X_i)\p{Y_i-\gamma_0(X_i)}}$. 
Thus, the RW estimator can be written as an infeasible ARW estimator using the true regression function. 

When balance is inexact, or when cross fitting is used, the RW and ARW estimators generally differ. In such cases, the ARW estimator is usually more stable because it also uses the estimated regression function. This point is consistent with the view that balancing and regression should be understood together rather than as competing ideas.

\paragraph{Cross fitting.}
Cross fitting is important in DML because it weakens empirical process conditions and helps justify the use of flexible nuisance estimators. It also changes how exact balance should be interpreted. If $\widehat\alpha$ is estimated on a training sample, exact balance on that training sample does not imply exact balance on the evaluation sample. Therefore, in cross-fitted DML, regressor imbalance should be assessed on the sample where the score is evaluated. This point is separate from the asymptotic role of cross fitting: cross fitting can justify the use of flexible nuisance estimators, while the remaining imbalance on the evaluation sample describes finite sample score error.

\paragraph{TMLE.}
TMLE and regressor balancing modify different nuisance components. TMLE updates the regression function so that an estimating equation is closer to being satisfied. Regressor balancing estimates the Riesz representer so that the empirical Riesz equation is better satisfied. Both approaches are motivated by the same Neyman orthogonal score, but they operate on different parts of the nuisance parameter. This distinction is useful for understanding when RW, ARW, and TMLE can be close to each other. If exact balancing holds for the relevant regression component, the RW estimator can be written in a form close to the ARW estimator. If balance is inexact, or if the regression and Riesz representer are estimated on different samples, the ARW estimator or TMLE can be more stable.

\paragraph{Related work.}
This study builds on several lines of work. \citet{Hainmueller2012entropybalancing} proposes entropy balancing, which constructs weights that exactly match prespecified covariate moments. \citet{Zubizarreta2015stableweights} proposes stable balancing weights, which minimize weight variability subject to approximate covariate balance constraints. \citet{Imai2013covariatebalancing} proposes CBPS, which estimates the propensity score by using covariate balancing moment conditions. \citet{Zhao2019covariatebalancing} proposes tailored loss functions and shows that the choice of loss should depend on the estimand and the link function. \citet{Fan2021optimalcovariate} studies how to choose covariate balancing functions in CBPS and shows that the optimal choice depends on outcome regression components. \citet{BrunsSmith2025augmentedbalancing} shows that augmented balancing weights and linear regression can be numerically equivalent in linear spaces. \citet{Kato2026aunified} develops a unified framework for Riesz representer estimation and shows that suitable Riesz regression problems imply balancing equations. Also see \citet{Benmichael2021balancingact}. 

We summarize the relationship among representative methods through two questions: which functions are balanced and how weight variability is controlled. Entropy balancing balances prespecified functions of $Z$ under an entropy criterion, while stable balancing weights balance prespecified functions of $Z$ while controlling weight variability \citep{Hainmueller2012entropybalancing,Zubizarreta2015stableweights}. 
We argue that the main difference is not whether a method uses weights, a propensity score, or a regression adjustment. The main difference is which functions are balanced. If the balanced functions depend only on $Z$, the method is covariate balancing. If the balanced functions depend on the full regressor $X=\p{D,Z}$, the method is regressor balancing. The latter is more general and is directly tied to the Neyman error decomposition in \eqref{eq:neyman_decomp}. Appendix \ref{app:taxonomy_details} gives further details.

\section{Simulation Studies}
\label{sec:experiment}
The experiments illustrate the difference between covariate balancing and regressor balancing from the viewpoint of the Neyman error. They are not intended to show that one estimator uniformly dominates another. Instead, they show that the imbalance relevant to the score can differ from ordinary covariate imbalance. We use the squared loss. Here, we report the simulation without cross fitting. We describe the details of the experiments in Appendix~\ref{app:experiment_details}. Additional experiments with cross fitting and other losses are reported in Appendix~\ref{app:additional_experiments}. We also investigate the performance using semi-synthetic data in Appendix~\ref{appdx:emp}.

We generate observations $\cb{\p{D_i,Z_i,Y_i}}^n_{i=1}$ with $n=1200$ and repeat the experiment 100 times. The covariates are $Z_i\in\bbR^3$ and follow the standard normal distribution. The treatment is generated from $D_i\sim\mathrm{Bernoulli}\p{e_0(Z_i)}$, where $e_0(Z_i)=\mathrm{expit}\p{0.5Z_{i1}-0.4Z_{i2}+0.2\sin(Z_{i3})}$. 
Let $\varphi(Z_i)\in\bbR^{80}$ be the basis constructed from random Fourier features for a Gaussian kernel. The outcome is generated as $Y_i=\mu_0(Z_i)+D_i\tau(Z_i)+\varepsilon_i$, 
where $\mu_0(Z_i)=\varphi(Z_i)^\top\beta_0$, $\tau(Z_i)=\psi(Z_i)^\top\beta_\tau$, and $\varepsilon_i$ is independent noise with standard deviation $0.05$. The coefficient vectors $\beta_0$ and $\beta_\tau$ are fixed across replications. The target is the sample ATE $\theta_0=\frac{1}{n}\sum^n_{i=1}\tau(Z_i)$.

We compare two ways of estimating the Riesz representer. Covariate balancing uses basis functions depending only on $Z$. Regressor balancing uses the treatment-specific basis $\Phi(D,Z)=\p{D\psi(Z),(1-D)\psi(Z)}$, so that the same basis $\psi(Z)$ is used and only the coefficients vary with treatment. The Riesz regularization parameter is $0.01$ in Table~\ref{tab:main_simulation}. Table~\ref{tab:main_simulation} shows that regressor balancing substantially reduces RW RMSE and regressor imbalance. ARW RMSE changes only slightly because the regression adjustment already removes part of the outcome error. Regressor balancing reduces the imbalance of the treatment-specific basis functions. It also improves the RW estimator, which depends directly on the estimated Riesz representer. The improvement for ARW is smaller because ARW also uses the regression adjustment.

\begin{table}[t]
\centering
\caption{Simulation study with squared loss and no cross fitting.}
\label{tab:main_simulation}
\small
\resizebox{\linewidth}{!}{
\begin{tabular}{lrrrrr}
\toprule
Method & RA (RMSE) & RW (RMSE) & ARW (RMSE) & Covariate Imbalance & Regressor Imbalance \\
\midrule
covariate & 0.021936 & 0.043803 & 0.009267 & 0.004297 & 0.008811 \\
regressor & 0.021936 & 0.009031 & 0.009087 & 0.006932 & 0.003541 \\
\bottomrule
\end{tabular}
}
\end{table}

\begin{table}[t]
\centering
\caption{Regularization sensitivity with squared loss and no cross fitting.}
\label{tab:main_regularization}
\small
\resizebox{\linewidth}{!}{
\begin{tabular}{lrrrrrr}
\toprule
Method & Reg. parameter $\lambda$ & RA (RMSE) & RW (RMSE) & ARW (RMSE) & Covariate Imbalance & Regressor Imbalance \\
\midrule
Covariate & 0.000000 & 0.021936 & 0.097923 & 0.013411 & 0.000082 & 0.016772 \\
Covariate & 0.010000 & 0.021936 & 0.043803 & 0.009267 & 0.004297 & 0.008811 \\
Covariate & 0.100000 & 0.021936 & 0.028418 & 0.010451 & 0.012699 & 0.007004 \\
Regressor & 0.000000 & 0.021936 & 0.004506 & 0.004374 & 0.000636 & 0.000464 \\
Regressor & 0.010000 & 0.021936 & 0.009031 & 0.009087 & 0.006932 & 0.003541 \\
Regressor & 0.100000 & 0.021936 & 0.020461 & 0.010496 & 0.015090 & 0.007566 \\
\bottomrule
\end{tabular}
}
\end{table}

\begin{figure}[!ht]
\centering
\begin{tabular}{cc}
\includegraphics[width=.46\textwidth]{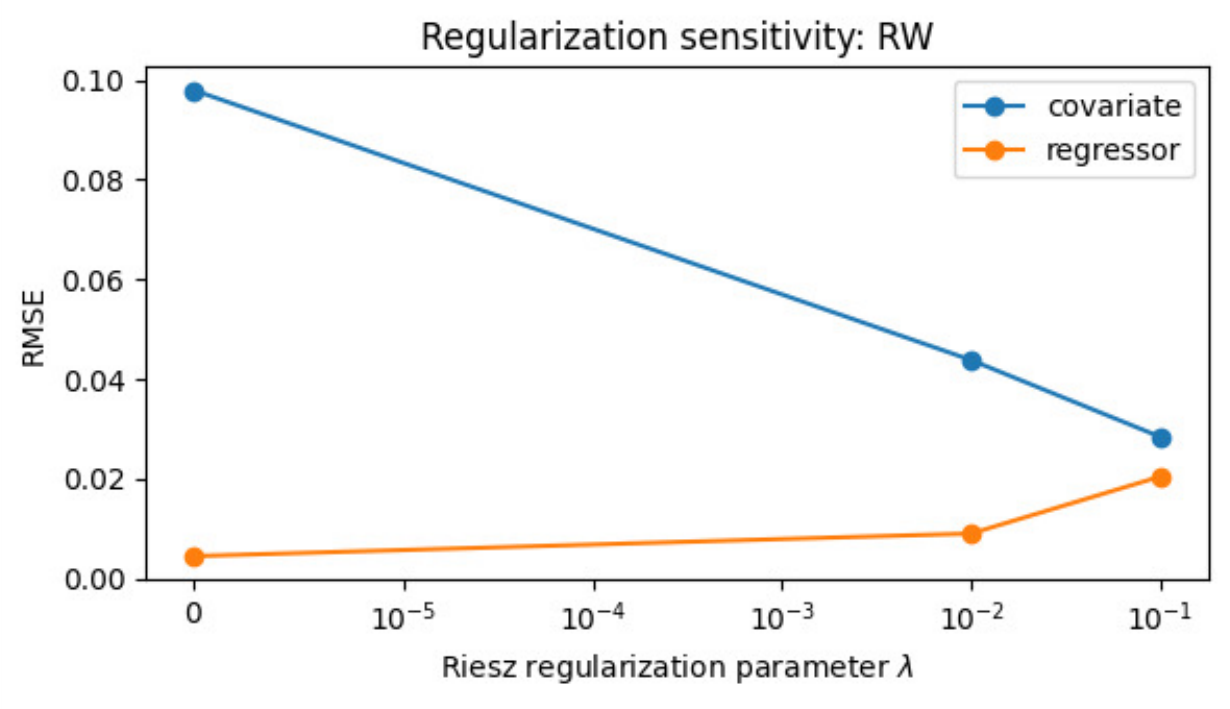} &
\includegraphics[width=.46\textwidth]{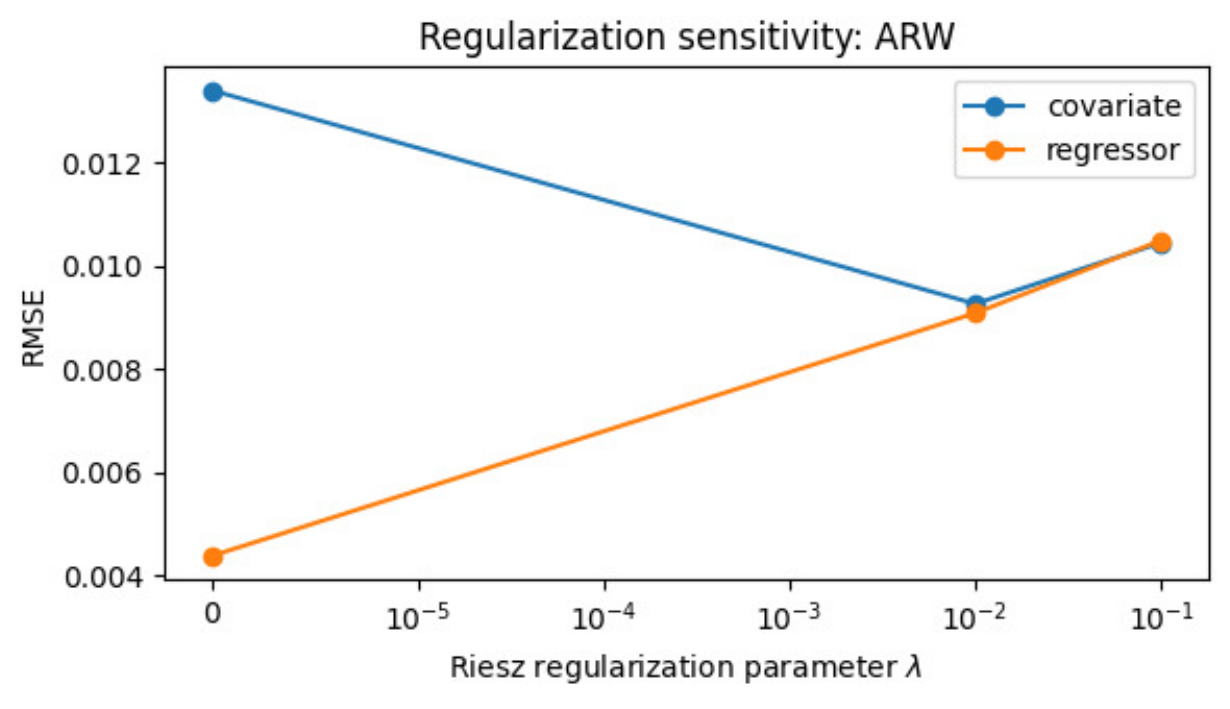} \\
\includegraphics[width=.46\textwidth]{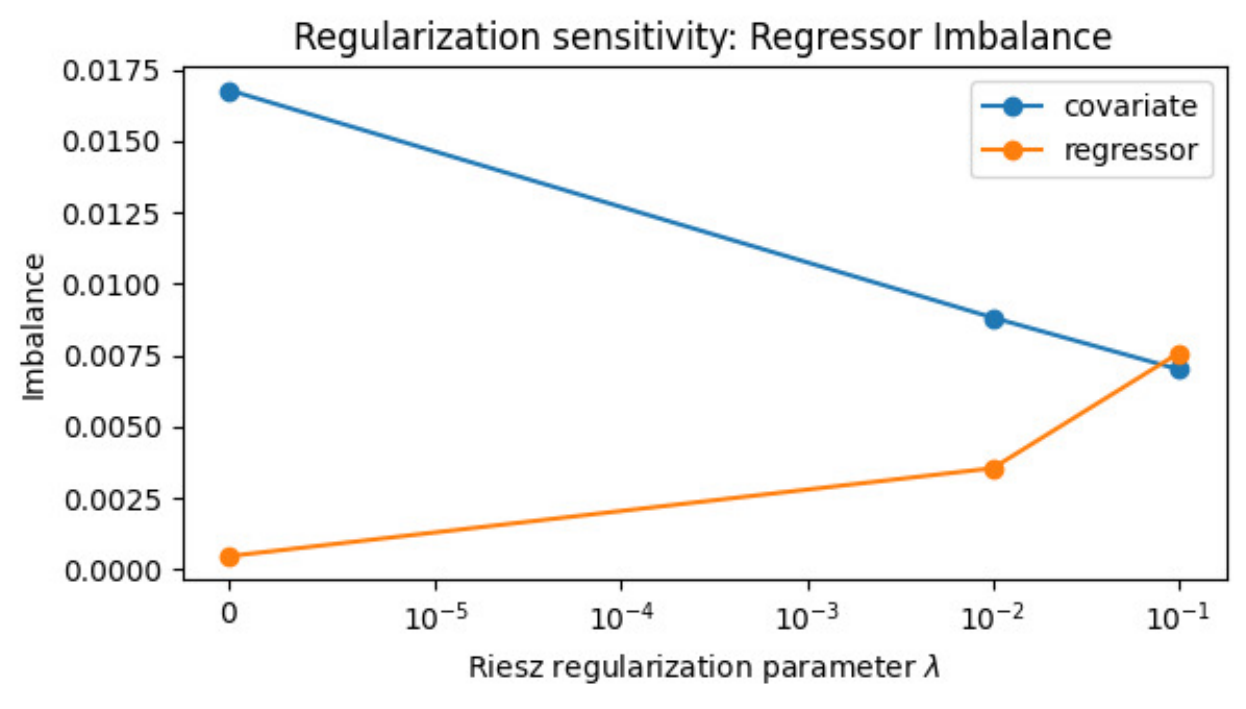} &
\includegraphics[width=.46\textwidth]{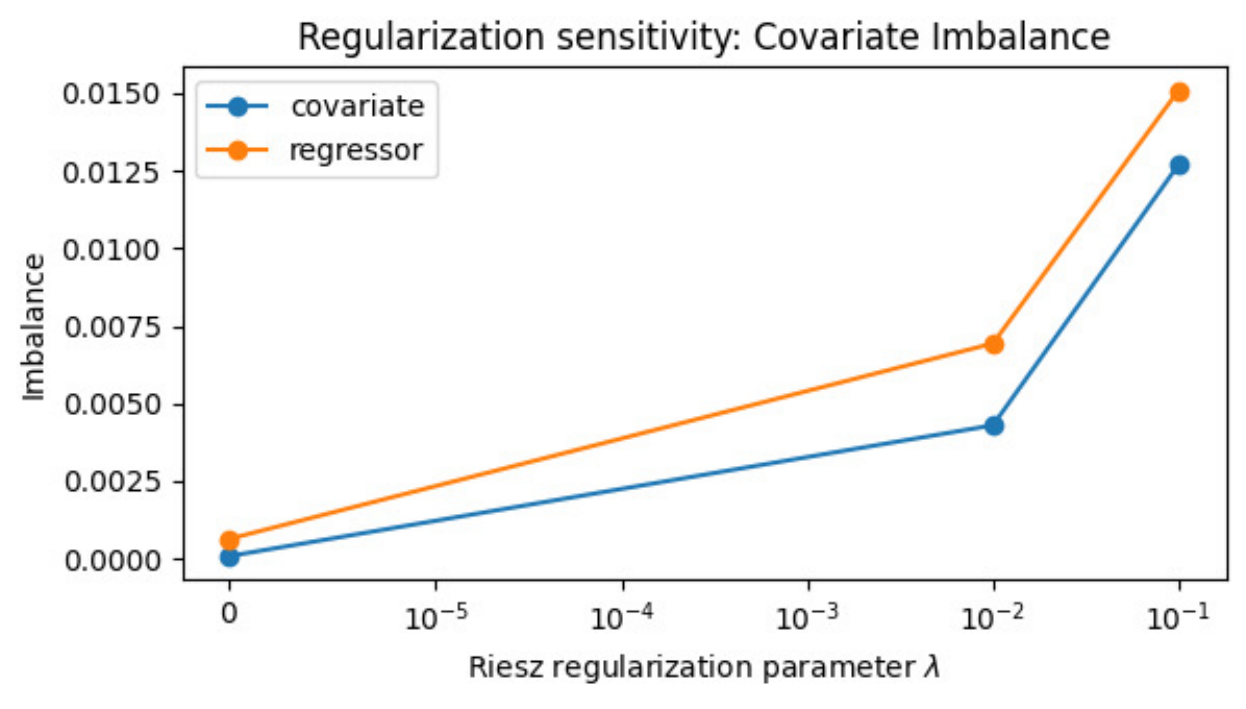}
\end{tabular}
\caption{Regularization sensitivity with squared loss.}
\label{fig:appendix_regularization}
\end{figure}

We also vary the Riesz regularization parameter over $0$, $0.01$, and $0.1$. Table~\ref{tab:main_regularization} and Figure~\ref{fig:appendix_regularization} show that the effect of regressor balancing depends on the regularization level. With $\lambda=0$, regressor balancing nearly eliminates regressor imbalance and gives the smallest RW and ARW RMSE in this design. Increasing $\lambda$ stabilizes Riesz regression but relaxes the balancing condition, and the regressor imbalance increases. The covariate balancing results show a different pattern. Very small regularization yields small covariate imbalance, but RW RMSE can be large. This supports the view that balance should be interpreted together with weight stability.

\section{Conclusion}
We reconsidered covariate balancing from the viewpoint of DML. The relevant object in DML is the Neyman orthogonal score, and the error of the estimated score depends on the regression components entering that score. Covariate balancing is useful when those components are functions of covariates alone, and it is sufficient in important cases such as ATT counterfactual mean estimation. However, for ATE under treatment effect heterogeneity, covariate balancing can be restrictive because the outcome regression depends on both treatment and covariates. Regressor balancing provides the more general condition: it balances basis functions of the full regressor and can remove treatment-dependent components of the score error. We therefore recommend reporting score-relevant regressor imbalance, not only covariate imbalance, when using balancing methods for estimating causal parameters.

\bibliography{arXiv2.bbl}

\bibliographystyle{tmlr}

\onecolumn

\appendix

\section{Alternative Views and Counterarguments}
\label{sec:alternative_views}

\paragraph{Covariate balance is enough for identification.}
One possible view is that, because unconfoundedness conditions on covariates, balancing functions of $Z$ should be sufficient. Our response is that identification and finite sample score error are different issues. Covariates identify the target under the causal assumptions, but the plug-in Neyman orthogonal score contains the regression error $\widehat\gamma(X)-\gamma_0(X)$. When this error depends on $X=\p{D,Z}$, balancing functions of $Z$ alone need not remove the deterministic part of the score error.

\paragraph{Propensity score weights already depend on treatment.}
Another view is that inverse propensity weighting already depends on treatment through the signed weight. Our point is not about whether the weight depends on $D$. It is about which functions are balanced. If the balanced functions are only $h(Z)$, then treatment-specific components such as $D h(Z)$ are not directly balanced.

\paragraph{Some ATE balancing methods already include treatment-specific moments.}
A third view is that some balancing methods for ATE already include moments that depend on treatment. We agree. In the terminology of this study, such methods are closer to regressor balancing than to covariate balancing alone. The distinction is therefore not a criticism of all weighting methods, but a recommendation to state which functions are balanced and whether they approximate the score-relevant regression error.

\paragraph{Exact balance can increase variance.}
A fourth view is that stronger balancing conditions can create unstable weights. We agree. Regressor balancing should not be interpreted as exact balance at any cost. The Neyman error decomposition contains both a deterministic imbalance term and a weighted noise term. Regularization and stable weights are therefore essential parts of the same principle.

\section{Additional Details for the Experiments}
\label{app:experiment_details}

This appendix gives additional details for the experiments in Section~\ref{sec:experiment}. The main text uses squared loss and no cross fitting. The covariate balancing and regressor balancing specifications use the same implementation, random seeds, feature construction, and optimization settings. All tables and figures are generated from the simulation design described below and from the IHDP semi-synthetic dataset described in Appendix~\ref{appdx:emp}.

\subsection{Data Generating Process}
In each replication, $Z_i$ is generated as a three-dimensional standard normal vector. The treatment assignment probability is
\begin{align}
e_0(Z_i)=\mathrm{expit}\p{0.5Z_{i1}-0.4Z_{i2}+0.2\sin(Z_{i3})},
\end{align}
and $D_i$ is drawn from $\mathrm{Bernoulli}\p{e_0(Z_i)}$. The outcome regression is generated from a Gaussian kernel feature map approximated by random Fourier features. Specifically, $\mu_0(Z_i)=\psi(Z_i)^\top\beta_0$ and $\tau(Z_i)=\psi(Z_i)^\top\beta_\tau$, where $\psi(Z_i)\in\bbR^{80}$. The observed outcome is $Y_i=\mu_0(Z_i)+D_i\tau(Z_i)+\varepsilon_i$, where the noise scale is $0.05$. The target is $\theta_0=\frac{1}{n}\sum^n_{i=1}\tau(Z_i)$ in each replication.

The outcome regression used by the estimator is intentionally more restrictive than the data generating process. It uses the same flexible feature map for the baseline component, but it includes only a constant treatment effect. This design creates a setting where the heterogeneous component of the treatment effect remains in the score error. Therefore, the experiment directly targets the distinction between covariate balancing and regressor balancing.

\subsection{Basis Functions and Riesz Regression}
Covariate balancing uses basis functions depending only on $Z$. Regressor balancing uses the treatment-specific basis $\Phi(D,Z)=\p{D\psi(Z),(1-D)\psi(Z)}$. In both cases, $\psi(Z)$ is the same random Fourier feature map. Thus, the comparison changes the dependence of the basis on treatment, not the underlying covariate feature map.

The main experiments use the squared loss. For ATE, a purely $Z$-only squared loss specification can be degenerate because $m(W;h)=0$ for functions $h(Z)$. Therefore, the squared loss is centered at the randomized-assignment ATE representer. This keeps the loss squared while allowing the $Z$-only specification to represent the usual signed covariate balancing condition. The regressor balancing specification uses the same loss construction and the treatment-specific basis. This design makes the comparison focus on whether the balanced basis depends only on $Z$ or on the full regressor $X=\p{D,Z}$.

\subsection{Reported Quantities}
The tables report RMSE for RA, RW, and ARW estimators. They also report the remaining imbalance for covariate functions and treatment-specific regressor functions. The RMSE of RW is particularly informative because it depends directly on the estimated Riesz representer. The RMSE of ARW is typically smaller and less sensitive because the regression adjustment removes part of the outcome error. The imbalance measures are reported to connect the simulation results to the decomposition in \eqref{eq:neyman_decomp}.

\section{Additional Experimental Results}
\label{app:additional_experiments}

This appendix reports additional experiments that are not shown in the main text. The main text focuses on squared loss without cross fitting. Here we report the comparison between no cross fitting and cross fitting, the sensitivity to the regularization parameter, and the results for alternative losses.

\subsection{Cross Fitting}
\begin{figure}[t]
\centering
\begin{tabular}{cc}
\includegraphics[width=.47\textwidth]{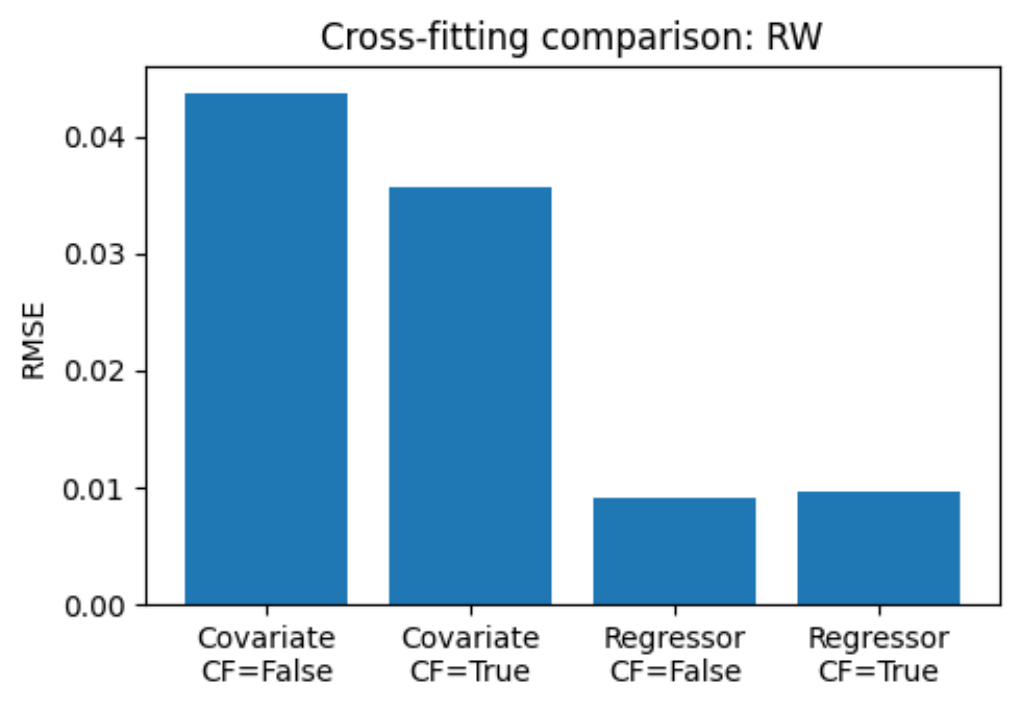} &
\includegraphics[width=.47\textwidth]{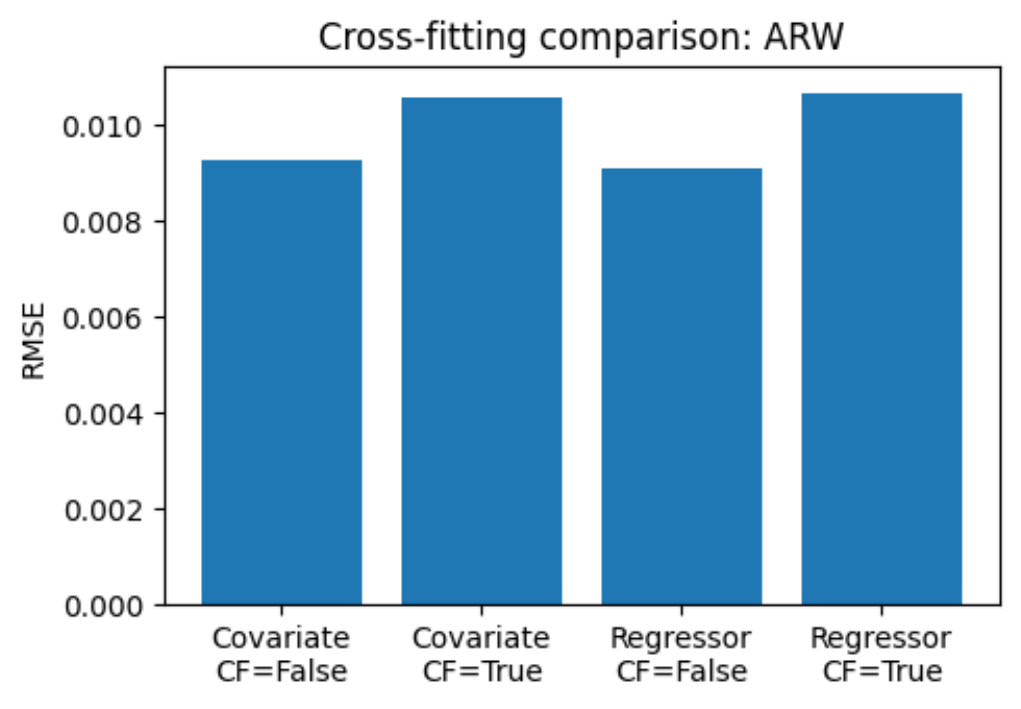} \\
\includegraphics[width=.47\textwidth]{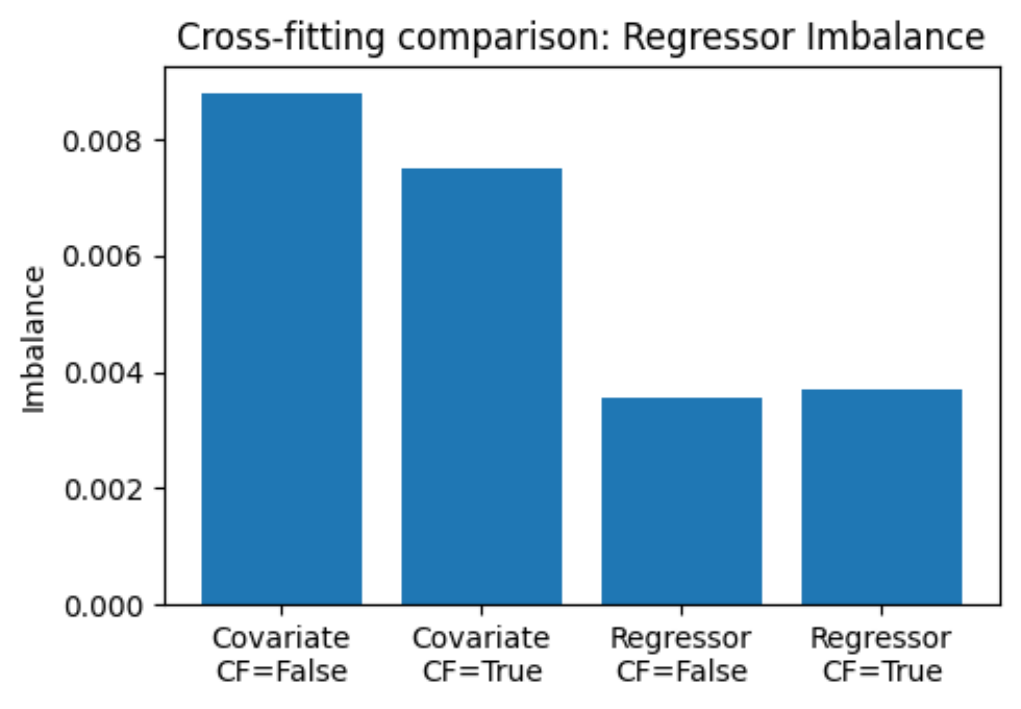} &
\includegraphics[width=.47\textwidth]{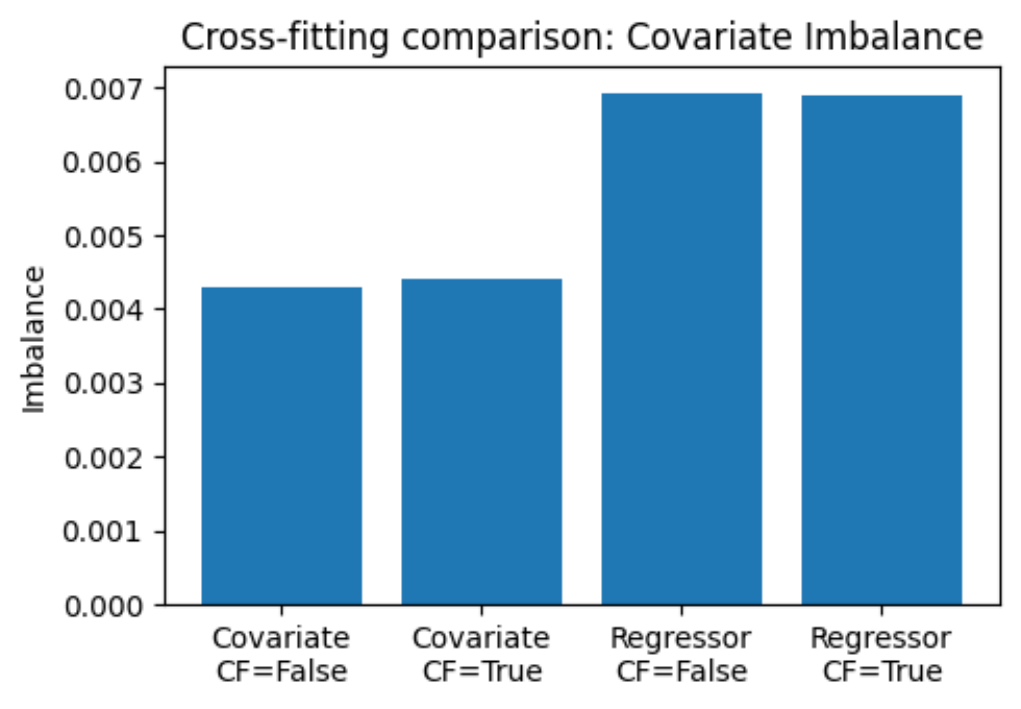}
\end{tabular}
\caption{Cross fitting comparison with squared loss.}
\label{fig:appendix_crossfit}
\end{figure}

\begin{figure}[t]
\centering
\includegraphics[width=.90\linewidth]{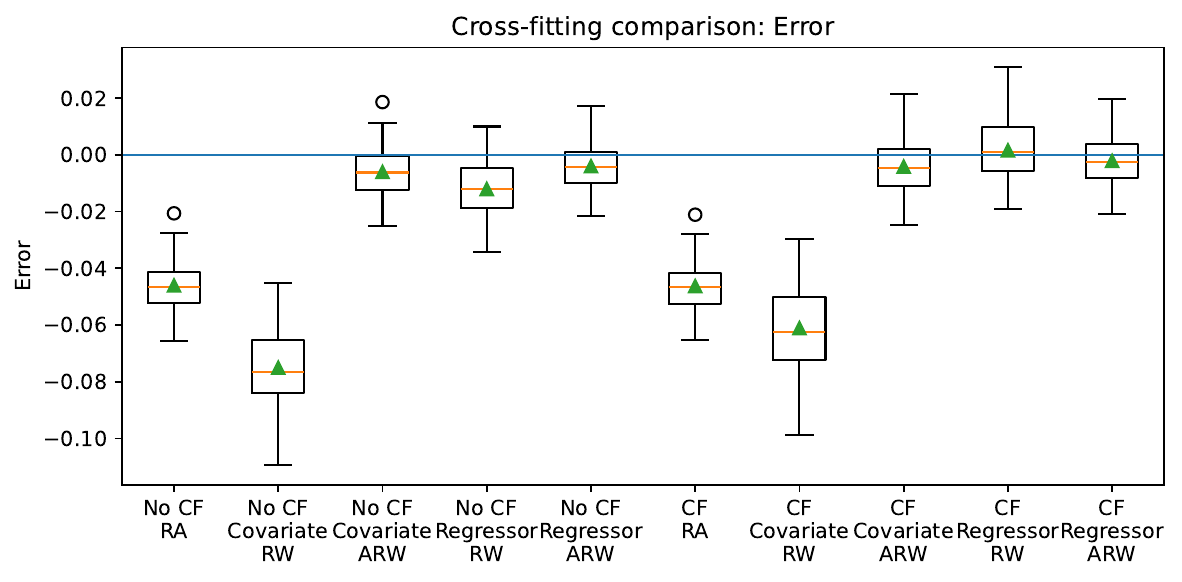}
\caption{Distribution of estimation errors in the cross fitting comparison.}
\label{fig:appendix_crossfit_boxplot}
\end{figure}

Table~\ref{tab:appendix_crossfit} and Figure~\ref{fig:appendix_crossfit} compare no cross fitting and cross fitting for squared loss. The fitting routine is the same in both cases, and the difference is whether the nuisance functions are evaluated on the same sample or on held-out folds. In this finite sample design, cross fitting slightly increases ARW RMSE. This does not contradict the role of cross fitting in DML. It reflects that exact balance on the training sample does not imply exact balance on the evaluation sample. Figure~\ref{fig:appendix_crossfit_boxplot} reports the distribution of estimation errors.

\begin{table}[t]
\centering
\caption{Cross fitting comparison with squared loss.}
\label{tab:appendix_crossfit}
\small
\resizebox{\linewidth}{!}{
\begin{tabular}{llrrrrr}
\toprule
Method & Cross Fit & RA (RMSE) & RW (RMSE) & ARW (RMSE) & Covariate Imbalance & Regressor Imbalance \\
\midrule
covariate & False & 0.021936 & 0.043803 & 0.009267 & 0.004297 & 0.008811 \\
covariate & True & 0.022284 & 0.035722 & 0.010566 & 0.004406 & 0.007509 \\
regressor & False & 0.021936 & 0.009031 & 0.009087 & 0.006932 & 0.003541 \\
regressor & True & 0.022284 & 0.009636 & 0.010671 & 0.006886 & 0.003688 \\
\bottomrule
\end{tabular}
}
\end{table}

\subsection{Alternative Losses}
The main text uses squared loss. Table~\ref{tab:appendix_losses} reports additional results for UKL and BP losses, together with the squared loss under the same alternative-loss experimental setting. The results show the same qualitative pattern. Regressor balancing reduces regressor imbalance and improves RW RMSE relative to covariate balancing in this design. ARW RMSE is less sensitive because it also uses the outcome regression. The loss names follow the terminology used in the Riesz regression implementation.

\begin{table}[t]
\centering
\caption{Alternative losses.}
\label{tab:appendix_losses}
\small
\resizebox{\linewidth}{!}{
\begin{tabular}{llrrrrr}
\toprule
Loss & Method & RA (RMSE) & RW (RMSE) & ARW (RMSE) & Covariate Imbalance & Regressor Imbalance \\
\midrule
bp & covariate & 0.048681 & 0.059602 & 0.011757 & 0.004780 & 0.007498 \\
bp & regressor & 0.048681 & 0.015065 & 0.010982 & 0.007169 & 0.003645 \\
sq & covariate & 0.048681 & 0.080690 & 0.012162 & 0.005252 & 0.010384 \\
sq & regressor & 0.048681 & 0.016793 & 0.011176 & 0.008373 & 0.004279 \\
ukl & covariate & 0.048681 & 0.029476 & 0.010773 & 0.003898 & 0.003630 \\
ukl & regressor & 0.048681 & 0.012795 & 0.010418 & 0.005685 & 0.002880 \\
\bottomrule
\end{tabular}
}
\end{table}

\section{Empirical Studies}
\label{appdx:emp}
We next use the IHDP semi-synthetic dataset. The dataset contains factual outcomes, treatment assignments, covariates, and the two conditional mean functions for each of 100 replications. Since both conditional mean functions are available, the true finite sample ATE can be computed in each replication. We combine the training and test files and compute the true ATE as the sample mean of $\mu_1(Z_i)-\mu_0(Z_i)$ in each replication. We then estimate the ATE from the factual outcomes in the same way as in the simulation study.

The empirical analysis uses squared loss Riesz regression without cross fitting. We again compare covariate balancing and regressor balancing. The basis functions are random Fourier features for an RBF kernel. Because the IHDP dataset has 25 covariates, we use the same number of features as in the simulation but set the kernel scale to 2.0. All other settings are the same as in the simulation study.

\begin{table}[t]
\centering
\caption{IHDP semi-synthetic study with squared loss and no cross fitting.}
\label{tab:ihdp_main}
\resizebox{\linewidth}{!}{
\begin{tabular}{lrrrrr}
\toprule
Method & RA (RMSE) & RW (RMSE) & ARW (RMSE) & Covariate Imbalance & Regressor Imbalance \\
\midrule
Covariate & 0.3252 & 0.4375 & 0.5075 & 0.0076 & 0.0098 \\
Regressor & 0.3252 & 0.3070 & 0.2966 & 0.0115 & 0.0071 \\
\bottomrule
\end{tabular}
}
\end{table}

Table~\ref{tab:ihdp_main} shows a pattern similar to that in the simulation study. Regressor balancing reduces regressor imbalance and improves RW and ARW RMSE relative to covariate balancing. In this experiment, covariate imbalance is larger under regressor balancing, so covariate imbalance alone does not explain the improvement. The improvement is instead aligned with the score-relevant regressor imbalance. This pattern is consistent with the Neyman error decomposition, although the result should be interpreted as illustrative rather than as a universal dominance claim.

\begin{figure}[t]
\centering
\begin{minipage}{0.48\linewidth}
\centering
\includegraphics[width=\linewidth]{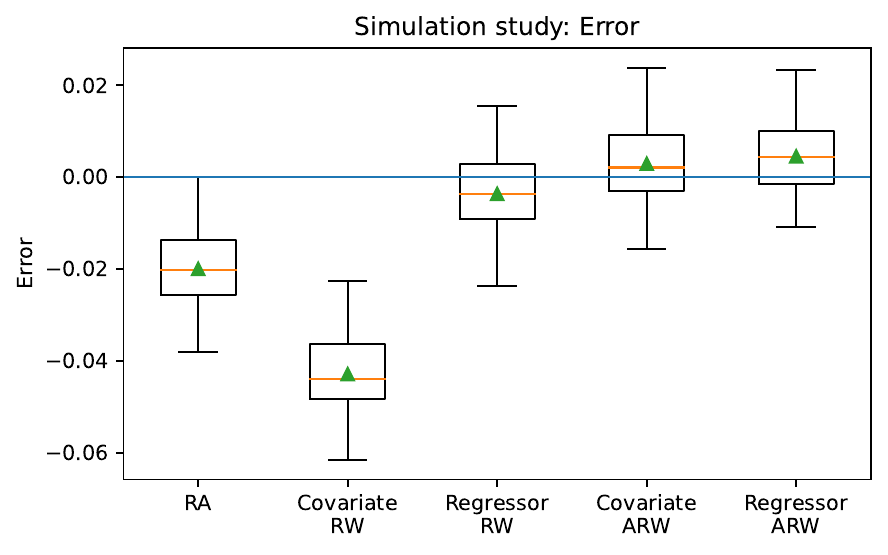}
\caption{Distribution of estimation errors in the simulation study.}
\label{fig:simulation_error_boxplot}
\end{minipage}
\hfill
\begin{minipage}{0.48\linewidth}
\centering
\includegraphics[width=\linewidth]{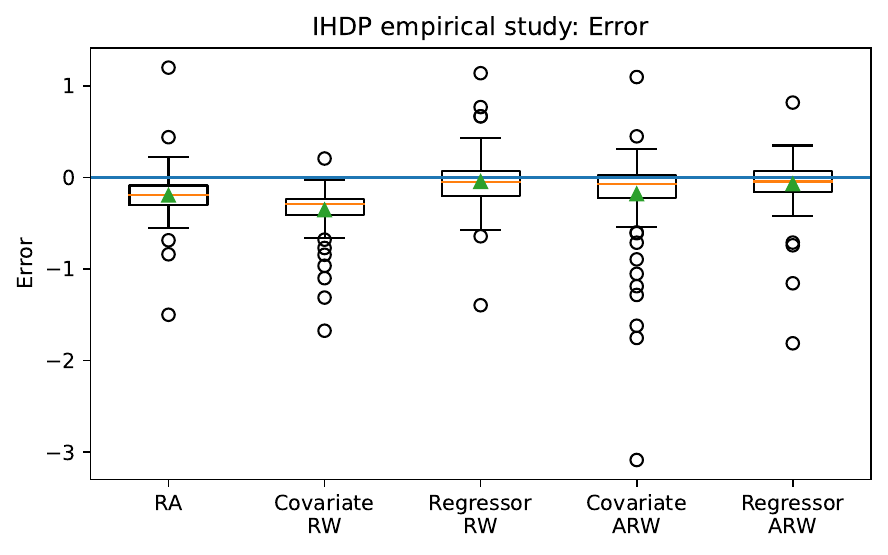}
\caption{Distribution of estimation errors in the IHDP semi-synthetic study.}
\label{fig:ihdp_error_boxplot}
\end{minipage}
\end{figure}

\section{Additional Details on Riesz Regression and Regressor Balancing}
\label{app:riesz_regressor_balancing}

\subsection{Riesz Regression and the Empirical Balancing Condition}
This appendix explains why Riesz regression naturally leads to regressor balancing. Recall that the Riesz representer $\alpha_0$ is defined by
\[
\bbE\sqb{m(W;\gamma)}=\bbE\bigsqb{\alpha_0(X)\gamma(X)}
  \quad\text{for all } \gamma\in L_2(P_X).
\]
The finite sample analogue of this identity is
\[
\frac{1}{n}\sum^n_{i=1}\alpha(X_i)f(X_i)
=
\frac{1}{n}\sum^n_{i=1}m(W_i;f).
\]
This is exactly the condition $\Delta_n(\alpha,f)=0$ defined above. Therefore, balancing is not an additional requirement imposed after estimating $\alpha_0$. It is the empirical version of the Riesz identity itself.

To see how Riesz regression gives this condition, suppose that we estimate $\alpha_0$ by a linear model $\alpha_\beta(X)=\beta^\top\Phi(X)$. Consider the squared loss Riesz regression objective
\[
\frac{1}{n}\sum^n_{i=1}\alpha_\beta(X_i)^2
-
\frac{2}{n}\sum^n_{i=1}m(W_i;\alpha_\beta)
+
\lambda J(\beta).
\]
Let $\widehat\beta$ be its minimizer and let $\widehat\alpha(X)=\alpha_{\widehat\beta}(X)$. If $\Phi_j$ denotes the $j$th component of $\Phi$, the first-order condition gives
\[
0
=
\frac{2}{n}\sum^n_{i=1}\Phi_j(X_i)\widehat\alpha(X_i)
-
\frac{2}{n}\sum^n_{i=1}m(W_i;\Phi_j)
+
\lambda \partial_j J(\widehat\beta).
\]
Equivalently,
\[
\Delta_n(\widehat\alpha,\Phi_j)
=
-\frac{\lambda}{2}\partial_j J(\widehat\beta).
\]
Thus, when $\lambda=0$, Riesz regression exactly balances the basis functions $\Phi_j$. When $\lambda>0$, Riesz regression approximately balances them, with the approximation determined by the regularization term. This is the sense in which Riesz regression automatically produces regressor balancing.

The important point is that the functions being balanced are the basis functions used to approximate functions of the full regressor $X$. In treatment effect problems, $X=\p{D,Z}$. Therefore, if $\Phi(X)$ depends on both $D$ and $Z$, Riesz regression balances functions of both treatment and covariates. If $\Phi(X)$ depends only on $Z$, the condition reduces to covariate balancing.

\subsection{Connection with Neyman Error}
The decomposition in equation \eqref{eq:neyman_decomp} shows that
\[
  \mathrm{NE}_n(\widehat\gamma,\widehat\alpha)
  =
  \frac{1}{n}\sum^n_{i=1}\widehat\alpha(X_i)\varepsilon_i
  -
  \Delta_n\p{\widehat\alpha,\widehat\gamma-\gamma_0}.
\]
The first term is a weighted noise term. Under cross fitting, conditional on the training data used to construct $\widehat\alpha$, this term has expectation zero. The second term is the deterministic part of the score error. Therefore, if $\widehat\gamma-
\gamma_0$ is well approximated by the basis functions that are balanced by $\widehat\alpha$, the deterministic part becomes small.

For example, suppose that
\[
\widehat\gamma(X)-\gamma_0(X)
=
\rho^\top\Phi(X)+r(X)
\]
for some coefficient vector $\rho$ and an approximation error $r$. Then, by linearity,
\[
\Delta_n\p{\widehat\alpha,\widehat\gamma-
\gamma_0}
=
\sum_j \rho_j \Delta_n(\widehat\alpha,\Phi_j)+\Delta_n(\widehat\alpha,r).
\]
If $\Delta_n(\widehat\alpha,\Phi_j)=0$ for every $j$ and $\Delta_n(\widehat\alpha,r)$ is small, then
\begin{align*}
\Delta_n\p{\widehat\alpha,\widehat\gamma-\gamma_0}
\end{align*}
is small. If the balancing conditions hold approximately, the same display shows that the deterministic part of the score error is controlled by the size of the coefficient vector, the remaining imbalance of the basis functions, and the approximation error. Thus, Riesz regression is useful not only because it estimates $\alpha_0$, but also because its first-order condition targets the balancing gap in \eqref{eq:neyman_decomp}.

\subsection{Approximate Balance and Weight Stability}
Regressor balancing should not be interpreted as requiring exact balance regardless of weight variability. The decomposition in \eqref{eq:neyman_decomp} contains a balance term and a weighted noise term. If $\widehat\alpha$ is too variable, the first term can be large even when the second term is small.

Assume that $\bbE\sqb{\varepsilon_i\mid X_i}=0$ and $\bbE\sqb{\varepsilon_i^2\mid X_i}\leq \sigma^2$. Conditional on $X_1,\ldots,X_n$ and on the data used to construct $\widehat\alpha$, we have
\[
\bbE\sqb{\p{\frac{1}{n}\sum^n_{i=1}\widehat\alpha(X_i)\varepsilon_i}^2\mid X_1,\ldots,X_n}
\leq
\frac{\sigma^2}{n^2}\sum^n_{i=1}\widehat\alpha(X_i)^2.
\]
Thus, reducing the Neyman error requires both small imbalance and stable weights. This explains why stable balancing weights and regularized Riesz regression are closely related to our motivation. The role of regularization is not only to make the numerical problem well posed. It also controls the variance contribution of the weighted noise term.

\section{ATE, Covariate Balancing, and Regressor Balancing}
\label{app:ate_covariate_regressor}

\subsection{ATE Riesz Equation}
For ATE estimation,
\[
m(W;\gamma)=\gamma(1,Z)-\gamma(0,Z).
\]
Therefore, for any function $f(d,z)$,
\[
m(W;f)=f(1,Z)-f(0,Z).
\]
The empirical Riesz balancing condition is
\[
\frac{1}{n}\sum^n_{i=1}\widehat\alpha(D_i,Z_i)f(D_i,Z_i)
=
\frac{1}{n}\sum^n_{i=1}\p{f(1,Z_i)-f(0,Z_i)}.
\]
This condition is regressor balancing because it is imposed on functions of the full regressor $X=\p{D,Z}$.

\subsection{Covariate Balancing as a Restricted Case}
Covariate balancing is obtained by restricting $f(d,z)$ to functions that do not depend on $d$. Let $f(d,z)=h(z)$. Then,
\[
m(W;f)=h(Z)-h(Z)=0.
\]
The empirical Riesz balancing condition becomes
\[
\frac{1}{n}\sum^n_{i=1}\widehat\alpha(D_i,Z_i)h(Z_i)=0.
\]
For the ATE Riesz representer,
\[
\alpha_0(D,Z)=\frac{\mathbbm{1}\p{D=1}}{e_0(Z)}-\frac{\mathbbm{1}\p{D=0}}{1-e_0(Z)},
\]
this condition is the usual signed covariate balancing condition. It balances treated and control groups after weighting in terms of the common function $h(Z)$.

This derivation clarifies why covariate balancing is a restricted case of regressor balancing. It is obtained by applying the Riesz equation only to functions of $Z$. This restriction is sufficient if the relevant regression error is represented by functions of $Z$ alone. It is restrictive if the relevant regression error depends on both $D$ and $Z$.

\subsection{Treatment-Effect Heterogeneity}
Write the outcome regression as
\[
\gamma_0(D,Z)=\gamma_0(0,Z)+D\p{\gamma_0(1,Z)-\gamma_0(0,Z)}.
\]
When the treatment effect is homogeneous, the difference $\gamma_0(1,Z)-\gamma_0(0,Z)$ is constant. In such a case, balancing common functions of $Z$ can remove the relevant part of the score error, provided that the regression error is represented by those functions.

When the treatment effect is heterogeneous, the difference $\gamma_0(1,Z)-\gamma_0(0,Z)$ depends on $Z$. Then, the regression error may also depend on both $D$ and $Z$. In that case, a basis function of the form $\widetilde{\Phi}(Z)$ may not represent the error $\widehat\gamma(D,Z)-\gamma_0(D,Z)$. A basis function $\Phi(D,Z)$ is then needed. This is the main reason why regressor balancing is more general than covariate balancing for ATE estimation.

\section{ATT and the Sufficiency of Covariate Balancing}
\label{app:att_covariate_balancing}
We do not claim that covariate balancing is wrong. ATT counterfactual mean estimation gives an important case where covariate balancing is natural.

Consider the counterfactual mean
\[
\bbE\bigsqb{\gamma_0(0,Z)\mid D=1}.
\]
The function to be transported from the control group to the treated group is $\gamma_0(0,Z)$, which is a function of $Z$ alone. Therefore, if $\gamma_0(0,Z)$ is well approximated by a basis $\widetilde{\Phi}(Z)$, then balancing $\widetilde{\Phi}(Z)$ between the treated group and the weighted control group directly targets this counterfactual mean.

This is why entropy balancing and related methods are particularly natural for ATT counterfactual mean problems. They choose weights for the control group so that specified functions of $Z$ match the corresponding moments in the treated group. If the specified functions of $Z$ approximate $\gamma_0(0,Z)$ well, then this balancing condition directly reduces the bias of the counterfactual mean estimator.

Thus, the distinction between covariate balancing and regressor balancing depends on the target estimand. For ATT counterfactual means, the regression function used in the score is $\gamma_0(0,Z)$, which is a function of $Z$ alone. For ATE under treatment effect heterogeneity, the outcome regression is generally a function of $X=\p{D,Z}$. This is why covariate balancing can be sufficient in the former case but restrictive in the latter case.

\section{Relation to CBPS and Optimal CBPS}
\label{app:cbps_ocbps}

\subsection{CBPS}
CBPS estimates the propensity score by imposing covariate balancing moment conditions. For ATE estimation, a typical balancing condition has the form
\[
\frac{1}{n}\sum^n_{i=1}
\p{
\frac{D_i}{e_\beta(Z_i)}
-
\frac{1-D_i}{1-e_\beta(Z_i)}
}
h(Z_i)
=
0.
\]
This is the same structure as the covariate balancing condition obtained by setting $f(d,z)=h(z)$ in the ATE Riesz equation. Therefore, standard CBPS can be interpreted as a method that estimates the propensity score by imposing a restricted version of regressor balancing, where the balanced functions depend only on $Z$.

This interpretation is consistent with the goal of CBPS. CBPS improves propensity score estimation by optimizing covariate balance rather than only treatment prediction. However, the balanced functions are usually functions of covariates. Therefore, standard CBPS does not automatically balance treatment-dependent basis functions unless such functions are explicitly included.

\subsection{Optimal CBPS}
Optimal CBPS clarifies that the choice of balancing functions matters. Let
\[
K(Z)=\bbE\sqb{Y(0)\mid Z},
\]
and
\[
L(Z)=\bbE\sqb{Y(1)-Y(0)\mid Z}.
\]
Optimal CBPS shows that balancing functions should be chosen so that they represent components related to $K(Z)$ and $L(Z)$. In this sense, optimal CBPS is closer to the regressor balancing perspective than standard CBPS.

A useful way to see this is through the two types of estimating functions used in optimal CBPS. One type has the form
\[
\p{
\frac{D}{e_\beta(Z)}
-
\frac{1-D}{1-e_\beta(Z)}
}
h_1(Z),
\]
which balances common components of the outcome regression. Another type has the form
\[
\p{
\frac{D}{e_\beta(Z)}
-
1
}
h_2(Z),
\]
which targets the treatment effect component. If $K(Z)$ is approximated by $h_1(Z)$ and $L(Z)$ is approximated by $h_2(Z)$, then these estimating functions target the components that enter the ATE score error.

Thus, optimal CBPS supports our main claim. The important question is not only whether covariates are balanced, but which functions should be balanced. For DML, this choice should be guided by the regression error entering the Neyman orthogonal score.

\section{Relation to Entropy Balancing and Stable Balancing Weights}
\label{app:entropy_stable}

\subsection{Entropy Balancing}
Entropy balancing constructs weights that exactly balance specified functions of covariates under an entropy criterion. In its common use for ATT estimation, the weights are assigned to the control group so that the weighted control group matches the treated group in specified covariate moments. This is a finite sample balancing method, and it is effective when the specified moments approximate the counterfactual regression function.

From our viewpoint, entropy balancing is a form of covariate balancing because the balanced functions usually depend only on $Z$. It becomes a form of regressor balancing when the target functional only requires balancing functions of $Z$, such as ATT counterfactual mean estimation. It remains restricted for ATE under treatment effect heterogeneity unless treatment-dependent functions are also balanced.

\subsection{Stable Balancing Weights}
Stable balancing weights choose weights that balance specified covariate functions while controlling weight variability. This is important because exact balance can lead to unstable weights in finite samples. The method therefore makes explicit the tradeoff between balance and stability.

From the viewpoint of Riesz regression, stable balancing weights are closely related to squared loss Riesz regression. Squared loss Riesz regression yields balancing equations through its first-order condition, and its dual form can be interpreted as a minimum variance balancing problem under appropriate specifications. Therefore, stable balancing weights can be viewed as a method that controls the same type of empirical Riesz imbalance while also controlling the variability of the weights.

This connection is important because regressor balancing should not be interpreted as requiring exact balance regardless of weight variability. The goal is to reduce the deterministic drift while keeping the weighted noise term stable. In the decomposition, we have
\[
\mathrm{NE}_n(\widehat\gamma,\widehat\alpha)=\frac{1}{n}\sum^n_{i=1}\widehat\alpha(X_i)\varepsilon_i-\Delta_n\p{\widehat\alpha,\widehat\gamma-\gamma_0},
\]
the second term is controlled by balance, while the first term is affected by the size and variability of $\widehat\alpha(X_i)$. Stable balancing weights address this tradeoff directly.

\section{Relation to Augmented Balancing Weights}
\label{app:augmented_balancing_weights}
Augmented balancing weights combine outcome regression and balancing weights. In linear settings, augmented balancing weights are closely related to linear regression. This relationship is important because it shows that balancing and regression are not separate principles. They can be two ways of expressing the same finite-dimensional approximation.

Suppose both the outcome regression and the balancing weights use the same basis functions. Then balancing those basis functions affects the same finite-dimensional space used by the regression estimator. If the basis functions depend only on $Z$, the method targets a covariate-only approximation. If the basis functions depend on $X=\p{D,Z}$, the method targets a richer regressor approximation.

This reinforces our main message. The key question is not whether one uses weighting or regression, but which basis functions are used. For DML, the basis functions should be chosen to approximate $\widehat\gamma(X)-\gamma_0(X)$ as it appears in \eqref{eq:neyman_decomp}. This is the reason for emphasizing regressor balancing rather than covariate balancing alone.

\section{Details for the Taxonomy}
\label{app:taxonomy_details}

This appendix gives additional details for the taxonomy in Section \ref{sec:discussion}. The purpose is not to introduce new methods, but to clarify how existing methods differ in the functions they balance and in the way they control weight variability.

\subsection{Entropy Balancing and Stable Balancing Weights}

Entropy balancing constructs weights so that prespecified functions of covariates are balanced exactly. In the common ATT setting, the weights are assigned to control observations so that the weighted control group matches the treated group in the specified functions of $Z$. For a function $h_j(Z)$, a typical balancing condition can be written as
\[
\frac{\frac{1}{n}\sum^n_{i=1}(1-D_i)\widehat w_i h_j\p{Z_i}}{\frac{1}{n}\sum^n_{i=1}(1-D_i)\widehat w_i}
=
\frac{\frac{1}{n}\sum^n_{i=1}D_i h_j\p{Z_i}}{\frac{1}{n}\sum^n_{i=1}D_i}.
\]
Thus, entropy balancing is a covariate balancing method when the functions $h_j$ depend only on $Z$. This is natural for ATT-type counterfactual means because the relevant counterfactual regression function, such as $\gamma_0\p{0,Z}$, is a function of $Z$.

Stable balancing weights also balance prespecified functions of $Z$, but they make the stability of the weights explicit. A simplified version of the idea is to choose weights with small variability subject to approximate balance constraints. For example, if $a_j$ denotes a target moment, the constraints can be written as
\[
\left|\frac{1}{n}\sum^n_{i=1}w_i h_j\p{Z_i}-a_j\right|\leq \delta_j.
\]
The exact optimization problem may include normalization and nonnegativity constraints, but the key point is that the method controls both balance and weight variability. This is important because reducing the imbalance term in \eqref{eq:neyman_decomp} is not sufficient by itself. The weighted noise term also depends on the size and variability of $\widehat\alpha\p{X_i}$.

\subsection{CBPS, Tailored Loss Methods, and Optimal CBPS}

CBPS estimates the propensity score by using moment equations that balance functions of $Z$. For ATE estimation, a typical condition has the form
\[
\frac{1}{n}\sum^n_{i=1}\p{\frac{D_i}{e_\beta\p{Z_i}}-\frac{1-D_i}{1-e_\beta\p{Z_i}}}h\p{Z_i}=0.
\]
This condition balances the function $h(Z)$ after inverse propensity weighting. Therefore, standard CBPS is naturally viewed as covariate balancing because the balanced function is a function of $Z$. The contribution of CBPS is that the propensity score is estimated while directly improving this balance, rather than only improving treatment prediction.

Tailored loss methods take a related but different route. They choose a loss function for propensity score estimation so that the first order condition implies covariate balance for the target estimand. In this sense, tailored loss methods connect the choice of loss function with the choice of balancing condition. In their standard use, the balanced functions are still functions used in the propensity score model, and these are usually functions of $Z$.

Optimal CBPS gives a sharper message about the choice of balancing functions. Write the outcome regression for ATE as
\[
\gamma_0\p{D,Z}=K\p{Z}+D L\p{Z},
\]
where $K(Z)$ is a baseline component and $L(Z)$ is a treatment effect component. Optimal CBPS uses two types of balancing functions. The first type has the form
\[
\frac{1}{n}\sum^n_{i=1}\p{\frac{D_i}{e_\beta\p{Z_i}}-\frac{1-D_i}{1-e_\beta\p{Z_i}}}h_1\p{Z_i}=0,
\]
and the second type has the form
\[
\frac{1}{n}\sum^n_{i=1}\p{\frac{D_i}{e_\beta\p{Z_i}}-1}h_2\p{Z_i}=0.
\]
The first condition is related to the baseline component $K(Z)$. The second condition is related to the treatment effect component $L(Z)$. Therefore, optimal CBPS supports our main claim. The choice of balancing functions should be guided by the regression components that enter the score error, not only by generic covariate moments.

\subsection{Riesz Regression and Augmented Balancing Weights}

Riesz regression estimates the Riesz representer directly. Suppose that the representer is modeled as $\alpha_\beta\p{X}=\beta^\top\Phi\p{X}$. A squared loss version of Riesz regression can be written as
\[
\min_{\beta}\cb{
\frac{1}{n}\sum^n_{i=1}\alpha_\beta\p{X_i}^2
-
\frac{2}{n}\sum^n_{i=1}m\p{W_i;\alpha_\beta}
+
\lambda J\p{\beta}
}.
\]
Let $\widehat\beta$ be a solution and define $\widehat\alpha\p{X}=\alpha_{\widehat\beta}\p{X}$. If $\Phi_j$ is the $j$th basis function, the first-order condition gives
\[
\Delta_n\p{\widehat\alpha,\Phi_j}
=
-\frac{\lambda}{2}\partial_jJ\p{\widehat\beta}.
\]
Thus, when $\lambda=0$, Riesz regression exactly balances the basis functions. When $\lambda>0$, it approximately balances them, with the remaining imbalance determined by the regularization term. This is the sense in which Riesz regression implements regressor balancing.

The important distinction is the argument of the basis functions. If $\Phi(X)$ depends only on $Z$, the balancing equations reduce to covariate balancing. If $\Phi(X)$ depends on $X=\p{D,Z}$, then the balancing equations are imposed on functions of both treatment and covariates. 

Augmented balancing weights further clarify the role of the basis. When the outcome model and the weighting model are both linear in a common basis, the resulting estimator can be written as a regression estimator using that basis. Therefore, the central issue is not whether the method is described as weighting or regression. The central issue is which basis functions are used. From the viewpoint of DML, those basis functions should approximate the regression error entering the Neyman orthogonal score.

\subsection{Connection to the Neyman Error}

The taxonomy above can be summarized through the Neyman error decomposition in \eqref{eq:neyman_decomp}. If the regression error can be written as
\[
\widehat\gamma(X)-\gamma_0(X)=\rho^\top\Phi(X),
\]
then linearity gives
\[
\Delta_n\p{\widehat\alpha,\widehat\gamma-\gamma_0}
=
\sum_j\rho_j\Delta_n\p{\widehat\alpha,\Phi_j}.
\]
Therefore, balancing the basis functions $\Phi_j$ directly controls the deterministic part of the Neyman error. If the basis functions depend only on $Z$, this argument applies only to regression errors represented by functions of $Z$. If the regression error depends on both treatment and covariates, then basis functions of $X=\p{D,Z}$ are needed.

This is the main distinction between covariate balancing and regressor balancing. Covariate balancing is sufficient when the relevant regression error is a function of $Z$. Regressor balancing is more general because it balances the basis functions used to approximate the full regression error in the Neyman score.

\end{document}